# Polarization Vortices in a Ferromagnetic Metal via Twistronics

Yingzhuo Lun[1,2,#,*], Xinxin Hu[2,3,#], Qi Ren[1,#], Umair Saeed[2,3], Kapil Gupta[2], Bernat Mundet[2], Ivan Pinto-Huguet[2,3], José Santiso[2], Jessica Padilla-Pantoja[2], José Manuel Caicedo Roque[2], Martin Børstad Eriksen[4,5], Gonzalo Merino[4,6], Yunpeng Ma[7], Qian Li[7], Bonan Zhu[1], Gang Tang[1], David Pesquera[2], Xueyun Wang[1], Jiawang Hong[1,8,*], Jordi Arbiol[2,9,*], and Gustau Catalan[2,9,*]

[1]School of Aerospace Engineering, Beijing Institute of Technology, Beijing 10081, China
[2]Catalan Institute of Nanoscience and Nanotechnology - ICN2 (CSIC & BIST), Barcelona, Catalonia 08193, Spain
[3]Autonomous University of Barcelona, Barcelona, Catalonia 08193, Spain
[4]Port d' Informació Científica (PIC), Campus UAB, 08193 Bellaterra (Barcelona), Catalonia 08193, Spain
[5]Institut de Física d'Altes Energies (IFAE), The Barcelona Institute of Science and Technology (BIST), Campus UAB, 08193 Bellaterra (Barcelona), Catalonia, Spain
[6]Centro de Investigaciones Energéticas, Medioambientales y Tecnológicas (CIEMAT), 28040 Madrid, Spain
[7]State Key Laboratory of New Ceramic Materials, School of Materials Science and Engineering, Tsinghua University, Beijing 10084, China
[8]Beijing Institute of Technology (Zhuhai), Zhuhai 519088, China
[9]Institució Catalana de Recerca i Estudis Avançats (ICREA), Barcelona 08010, Catalonia

[#]These authors contributed equally to this work.
[*]Corresponding author emails: lunyingzhuo@bit.edu.cn (Y.L.), hongjw@bit.edu.cn (J.H.), arbiol@icrea.cat (J.A.), gustau.catalan@icn2.cat (G.C.)

**Abstract**

Recent advances in moiré engineering have revealed new pathways for manipulating lattice distortions and electronic properties in low-dimensional materials via stacking two crystalline layers with a twist angle between them, aka “twistronics”. Here, we demonstrate that twisted bilayer stacking can induce dipolar vortices in metallic $SrRuO_3$ membranes, despite the presence of free charge. These vortices are correlated with moiré-periodic flexoelectricity induced by shear strain gradients, and their magnitude depends both on the twist angle and proximity to the interface. In addition, below the ferromagnetic Curie temperature of the films, ferromagnetic order coexists - and appears to compete - with the dipolar order, showing opposite twist-angle dependencies of the respective magnitudes of the two multiferroic order parameters. Density functional theory calculations provide insights into the microscopic origin of these observations. These findings extend the scope of topological polarization design beyond dielectric materials and into metals.

**Highlights:** Twistronics enables the emergence of polarization vortices and multiferroic characteristics in metals.

Stacking two thin crystalline layers rotated with respect to each other generates a moiré interference pattern in the electronic density of states at the interface, which in turn modifies the electronic properties of the bilayer system. Since the discovery of superconductivity in twisted bilayer graphene[1], the field of so-called "twistronics"[2,3] has blossomed. While twistronics was initially studied in two-dimensional (2D) materials[4], which can be readily exfoliated and stacked, the development of methods for growing epitaxial films of complex oxides and separating them from the substrate *via* dissolution of a sacrificial layer[5–7] also allows these materials to partake in the twistronic game. A seminal result in this context has been the discovery that twisted bilayers of $BaTiO_3$ display polarization vortices[8]. The crystal and electronic structures of such twisted interfaces were investigated[9,10], and the proposed explanation for such vortices is that the torsional strain introduced at the interface between the twisted layers is not homogeneous, being minimal at the sites where the atomic columns of the two layers coincide across the interface, and maximal where equivalent atomic columns of the two layers are furthest apart. This difference in bonding strength causes torsional strain gradients and therefore gradient-induced polarization, known as flexoelectricity[11,12].

Flexoelectricity is allowed in materials of any symmetry, and twist-induced polar vortices have also been reported in cubic $SrTiO_3$[13,14], which is not polar to start with - although strain can make it ferroelectric[15]. On the other hand, there are no strain gradients without strain, so it can be difficult to separate gradient-induced flexoelectricity from strain-induced ferroelectricity or piezoelectricity in materials prone to polarize under mechanical stress. While piezoelectricity has been ruled out from the explanation of vortices based on geometrical considerations, the fact remains that, at least in $BaTiO_3$ and possibly also in strained $SrTiO_3$, there is an initial ferroelectric polarization which can be affected by strain and/or by depolarization fields inherent in the interfacial disruption of polarization in dielectric materials. In addition, the disruption of the vertical Ti-O-Ti chains at the interface is also expected to create local electrostatic fields which can cause a local dielectric polarization. It is therefore desirable to examine whether polar vortices can be induced in a material where piezoelectricity and interfacial electrostatic fields can be fully excluded. In turn, confirming the flexoelectric origin of polar vortices would suggest that the search for chiral polarization[16,17] need not be confined to dielectrics. Since metals can also be flexoelectric[18–20], a tantalizing possibility emerges of inducing polarization vortices in a metal.

In this work, we have fabricated twisted-bilayer (t-BL) stacks of $SrRuO_3$ (SRO), a well-known oxide metal whose flexoelectricity has been reported[19], and we have combined atomic-scale imaging and 3D ptychography atomic scale multislice reconstructions with theoretical calculations to examine the polar displacement pattern. The results show that the Ru cation is shifted with respect to the Sr lattice, forming periodic arrays of vortices and antivortices like those reported in insulating $BaTiO_3$ and $SrTiO_3$[8,13], displaying twist dependence and penetration depth. Moreover, since SRO is ferromagnetic below a Curie temperature of *c.a.* 140 K, at low temperatures the twisted bilayer membranes display a multiferroic coexistence of polar and magnetic ordering. The results further indicate that the two ferroic orders are anticorrelated, suggesting magnetoelectric coupling, with measurable consequences for the electronic transport properties.

**Film growth and structural characterization**

The starting epitaxial thin films of SRO (~10 nm) were grown by pulsed-laser deposition (PLD) on (001)-oriented $SrTiO_3$ (STO) substrates with water-soluble $Sr_3Al_2O_6$ (SAO) buffer layers. The films were attached onto a polydimethylsiloxane (PDMS) stamp and immersed in deionized water to dissolve the SAO layer. Upon removal of SAO and separation of SRO from the substrates, the freestanding single-layer (SL) SRO membranes were transferred onto silicon oxide or TEM supports. The t-BL SRO was assembled by stacking an additional membrane, released from the same epitaxial film and transferred onto the oxygen plasma-treated bottom layer with a prescribed rotational angle. To remove potential surface adsorbates at the interface and/or hydrogen intercalation[21], the transferred samples were annealed at 330 °C for 8 hours in an oxygen atmosphere. The schematic of the fabrication process is provided in Fig. 1a, and further details of film growth and transfer are described in Methods.

The crystallinity of the initial epitaxial films and subsequently transferred membranes was confirmed by X-ray diffraction (XRD) 2θ-ω scans (Fig. 1b). The fully epitaxial strained state of the as-grown heterostructures was confirmed using reciprocal space mapping (Fig. 1c), revealing that the SRO and SAO layers share the same in-plane lattice parameter as the STO substrate. The diffraction peak of the transferred membrane (Fig. 1b), meanwhile, is shifted, indicating a relaxation of the lattice upon epitaxial strain release. The as-grown film and transferred bilayer membranes (Supplementary Fig. S1) show smooth and flat surfaces with roughness of 0.25 nm and 0.38 nm, respectively. The bilayer interfacial quality was examined through cross-sectional scanning transmission electron microscopy with high-angle annular dark-field (STEM-HAADF) imaging (Fig. 1d and Extended Data Fig. 1). The intensity of atomic columns decays towards the interface, yielding a dark-contrast region where the atoms are only faintly observed. Tilting both layers with respect to the electron beam's zone axis (Fig. 1d) shows well defined atomic planes, with an uneven interfacial gap, alternating $RuO_2$-terminated separations of ~6.69 Å with SrO-terminated structural bridges, as highlighted by the white dash circles, where the interlayer separation is only ~3.1 Å. We expect the bonding across such regions to be strong enough to enable mechanical coupling across interface, and this expectation is confirmed by strain mappings in following planar-view STEM measurements. We also point out that higher temperature annealing yield tighter interfaces with periodic arrays of dislocations[22]. While at distances comparable to the periodicity of the structural singularities the differences may add complexity to the local polar texture, far away from the interface the strain gradients get smoothed out and are well described by continuum elasticity.

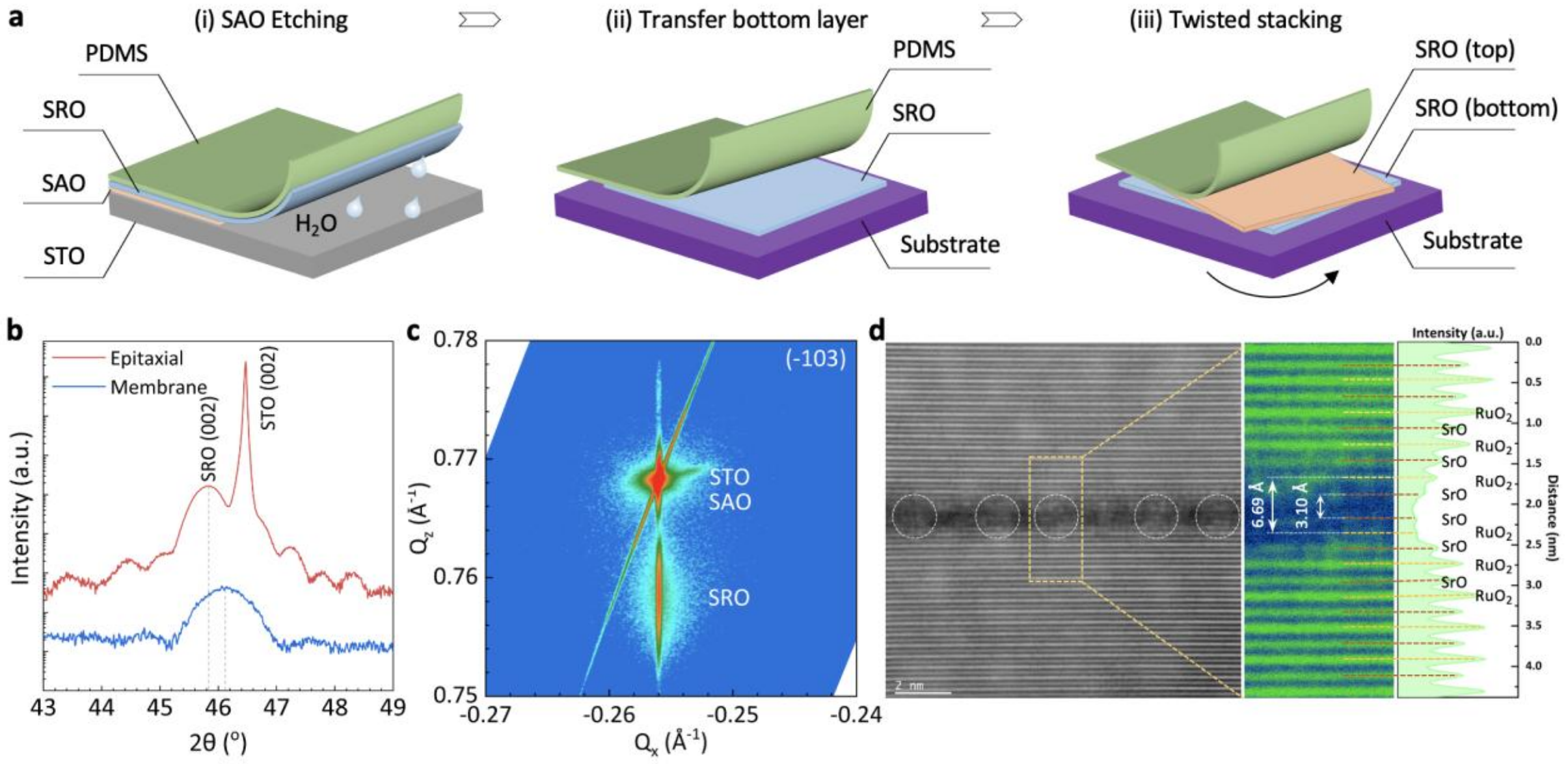

**Figure 1. Transfer and characterization of twisted-bilayer $SrRuO_3$ membranes.** **a**, Schematic illustration of the transfer process of twisted bilayer membranes. **b**, X-ray diffraction 2θ-ω scans of the as-grown epitaxial film and transferred membrane (a.u., arbitrary units). **c**, Reciprocal space mapping of the as-grown film around the (−103) diffraction peak. **d**, Atomic resolution cross-sectional HAADF-STEM characterization of the interface of a 4.8° twisted-bilayer with misaligned zone axis. The white dashed cycles in the left panel highlight the small-gap structural connections. The right panel shows a magnified area at around the interface, with corresponding intensity line profile showing the position of the (001) atomic planes.

## Polarization vortices in twisted-bilayer $SrRuO_3$

To visualize the lattice structure and depth-resolved dipole evolution in planar view, we used both high-resolution STEM-HAADF imaging and multislice electron ptychographic STEM (see Methods). First, the bulk-like orthorhombic symmetry (Pbnm)[23] of the SL was confirmed by the presence of ½ superlattice spots in the selected area electron diffraction (SAED) pattern and antiferrodistortive octahedral rotation in ABF-STEM image (Supplementary Fig. S2). To map the structural dipoles, we measured the off-center displacement of Ru ($\delta_{Ru}$) with respect to its four adjacent Sr atoms. The Ru displacement extracted from the STEM-HAADF image of the SL SRO (Figs. 2a, e) is random and minimal (1.3 pm on average), indicating that the freestanding SRO remains nonpolar (although, in polydomain membranes, polarization can emerge around ferroelastic walls[24].

Meanwhile, the t-BL SRO exhibits a well-ordered interfacial moiré (Supplementary Fig. S3). High-resolution images measured from the 3.0°, 4.8°, and 7.3° twisted bilayers are shown in Figs. 2b-d, displaying a decreasing moiré periodicity with increasing twist angle. The twist angle was determined from the relative rotation between the two distinct sets of diffraction spots observed in the fast Fourier transform (FFT) of the moiré images (Supplementary Fig. S4). Subsequently, we shifted the focus away from the interface to the upper surface to selectively image the lattice structure of the top layer (Supplementary Fig. S5). We then conducted depth-dependent STEM imaging and analyzed the corresponding variations in FFT patterns, confirming that the diffraction spots from the bottom lattice vanish as the focus shifts to the upper surface (Supplementary Fig. S6). We are therefore confident that there are no geometric interference artifacts from the bottom layer in the images obtained at the top layer; we will return to this point later. The Ru displacement maps of the top layers (Figs. 2f-h), calculated based on the STEM-HAADF images (Supplementary Fig. S5), reveal a periodic arrangement of vortices in the metallic membranes.

The toroidal moment was quantified as $\mathbf{Q}_i = \frac{1}{2N}\sum_1^N \mathbf{r}_i \times \boldsymbol{\delta}_i$, where $\boldsymbol{\delta}_i$ is the adjacent Ru displacement vector located at position $\mathbf{r}_i$, and $N = 24$ is the number of adjacent cells, and then superimposed onto the Ru displacement maps to visualize the spatial arrangement of polar vortices (Figs. 2f-h). Herein, “AA” and “AB” stacking denote, respectively, Sr–Sr or Ru–Ru sites aligned vertically across the interface (AA), or Sr–Ru vertically aligned (AB), in analogy to the nomenclature used in twisted van der Waals materials[1] and adopted in other twisted perovskite bilayers[8]. The periodic array of negative toroidal moments reveals that the AA-stacked regions host clockwise vortices, while anticlockwise vortices form near the AB- stacked regions. These features are reproduced in all bilayers even with a high twist angle of 10.4° (Extended Data Fig. 2).

As the twist angle increases, the periodicity of the polar vortices becomes smaller, coinciding with the moiré periodicity. The magnitudes of the toroidal moment and the polarization also decrease as the twisting increases. Figure 2i shows that Ru displacements exhibit an inverse dependence on the twist angle, being bigger at smaller twist angles. The maximum displacement reaches ~20 pm (8.3 pm on average) for the 3.0° twisted bilayer, which is comparable to reported values for polarized SRO[19,25,26]. Geometric phase analysis (GPA) (see Supplementary Text I) shows shear strain gradients of alternating signs forming a strip network, with the polarization components pointing consistently within each strip (Extended Data Fig. 4); the flexoelectric fields reconstructed by shear strain gradients (Figs. 2j-l) correlate with the observed polar vortices maps. The strain gradients scale inversely with the twist angle, Supplementary Fig. S8, replicating the polarization behavior.

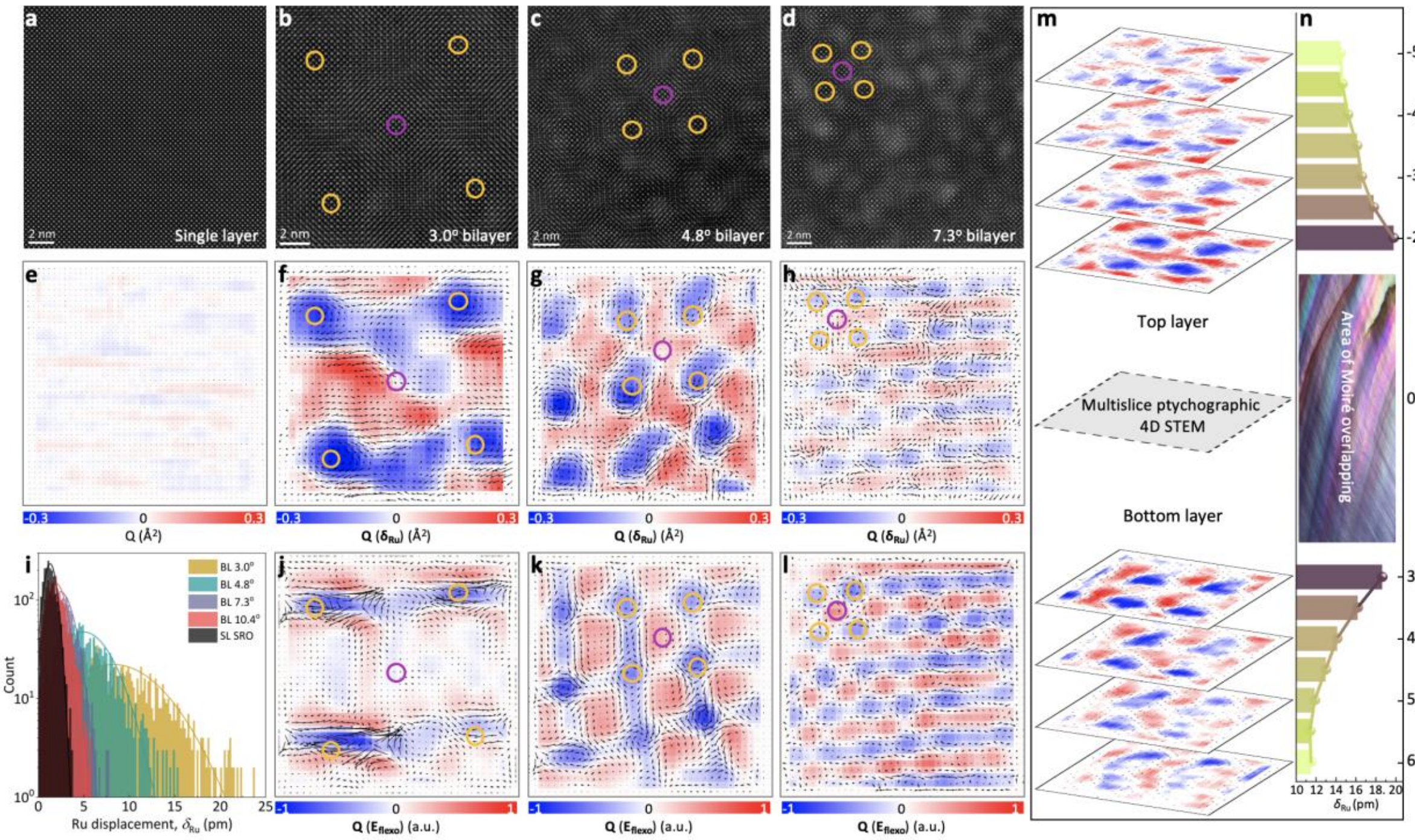

**Figure 2. Atomic-scale imaging of polar vortices in t-BL SROs.** **a**, Planar-view STEM-HAADF image of the freestanding SL SRO. **b-d**, Moiré fringes appear in HAADF images focusing on the interface in the 3.0°, 4.8°, 7.3° twisted bilayers, respectively. **e**, Ru displacement vector maps superimposed with their toroidal moment $Q(\delta_{Ru})$ for the SL SRO. **f-h**, Same Ru displacement analyses for the top layer in the 3.0°, 4.8°, 7.3° twisted bilayers, respectively. The Ru displacements in **e**, **f**, **g**, and **h** are extracted from HAADF images, and amplified by factors of 30, 30, 50, 100 for clarity, respectively. **i**, Histograms comparing the magnitude of Ru displacement of SL and t-BL SROs. **j-l**, Normalized flexoelectric field vector maps reconstructed by GPA-calculated shear strain gradients, superimposed with their toroidal moment $Q(E_{flexo})$. The yellow and purple circles in **b-d**, **f-h**, and **j-l** indicate the AA- and AB-stacked regions, respectively. **m**, Multislice ptychographic STEM reconstruction of depth-resolved polarization vortex. Depth denotes the distance from the interface. **n**, Corresponding mean Ru displacement (with respect to octahedral) of each slice image as a function of depth.

To exclude the possibility that the observed polarization vortices arise from moiré optical artifacts, we performed two additional experiments. First: we repeated the HAADF-STEM measurements on unannealed bilayers, the idea being that the unannealed interface could have adsorbates and/or increased interfacial gaps, resulting in a weaker interlayer coupling and lower polarization. Indeed, the unannealed samples remain non-polar, with a Ru displacement map (Extended Data Fig. 3) indistinguishable from that of SL SRO (Fig. 2e). Since the moiré exists both in the annealed and unannealed bilayers but the vortices are only observed in the annealed bilayers, the implication is that the vortices are not moiré interference artifact. This result is also consistent with a recent report that twisted bilayers with large interfacial gaps of the order of 1.5 nm[27] have no vortices. For the second test, we employed 4D STEM multislice electron ptychography[28–30] to reconstruct the depth-resolved lattice structure of a 4.8° twisted bilayer. The depth-resolved vortex distributions within the top and bottom layer (Fig. 2m) match the HAADF-STEM observation. In addition, the polarization vortices become progressively clearer. The depth-resolved mean Ru displacements in Fig. 2n show that the polarization magnitude increases toward the interface, supporting an interfacial origin (see also Extended Data Fig. 5 and Supplementary Movie S1). Quantitative analysis of the polarization profile as a function of depth further reveals an exponential decay with a characteristic length of ~7.5 nm (Supplementary Fig. S9), consistent with the strain relaxation length (~7.4 nm) predicted by continuum strain-gradient elasticity (Supplementary Text II). This agreement between polarization and strain relaxation length scales provides further support for flexoelectricity. In contrast, the Thomas–Fermi screening length of metallic SRO is sub-nanometric[27], precluding long range dipole-dipole interactions. All the available evidence is thus consistent with a flexoelectric origin of the vortices.

While polarization is hard to quantify or even define in a metal, we can nevertheless provide a nominal value using the atomic displacements and the ionic charge of the atoms. Dividing this nominal polarization by the strain gradient (see Supplementary Text III), we obtain a flexoelectric coefficient of 1.39 nC/m, which is of the same order of magnitude as typical values reported for centrosymmetric materials. Although this coefficient is only nominal – being based on nominal charges rather than Born effective charges (which are ill-defined in a metal), its plausible magnitude is reassuring.

**DFT calculation for polarized t-BL $SrRuO_3$**

To further probe our conclusions, we have used first-principles density functional theory (DFT) to simulate the polarization in t-BL SRO (see Methods). Commensurate bilayer supercells were constructed by stacking two individual unit-cell-thick layers with a relatively high twist angle of 18.92°; the smaller moiré periodicity with fewer atoms in a high-angle model improves computational efficiency. The model exhibits the characteristic moiré pattern arising from alternating AA- and AB-stacked sequences (Fig. 3a). After relaxing the structure, we computed the off-center displacement of the Ru atoms relative to the surrounding oxygen octahedra in each unit cell, revealing the formation of dipole vortices of opposite vorticity in the top and bottom layers (Figs. 3b-c). In the top layer, clockwise vortices form at AA-stacked regions and rough anticlockwise ones at AB-stacked regions, in agreement with experiment. Similar polarization configurations were reproduced in bilayers with higher twist angles of 22.62° and 28.07° (Supplementary Fig. S10). As in the experiments, the calculated Ru displacement decreases with twist angle (Fig. 3d), albeit the magnitudes are larger because the model bilayers are thinner and thus closer to the interface. In contrast, 0°-stacked bilayers with different translational shifts maintain a centrosymmetric structure without any polarization (Supplementary Fig. S11), evidencing the necessary role of twisting. Importantly, all twisted bilayers exhibit nonzero density of states at the Fermi level, primarily contributed by Ru *d*-orbital electrons with a strong Ru-O orbital hybridization, confirming their metallic nature. Topological polarization is thus theoretically compatible with metallicity.

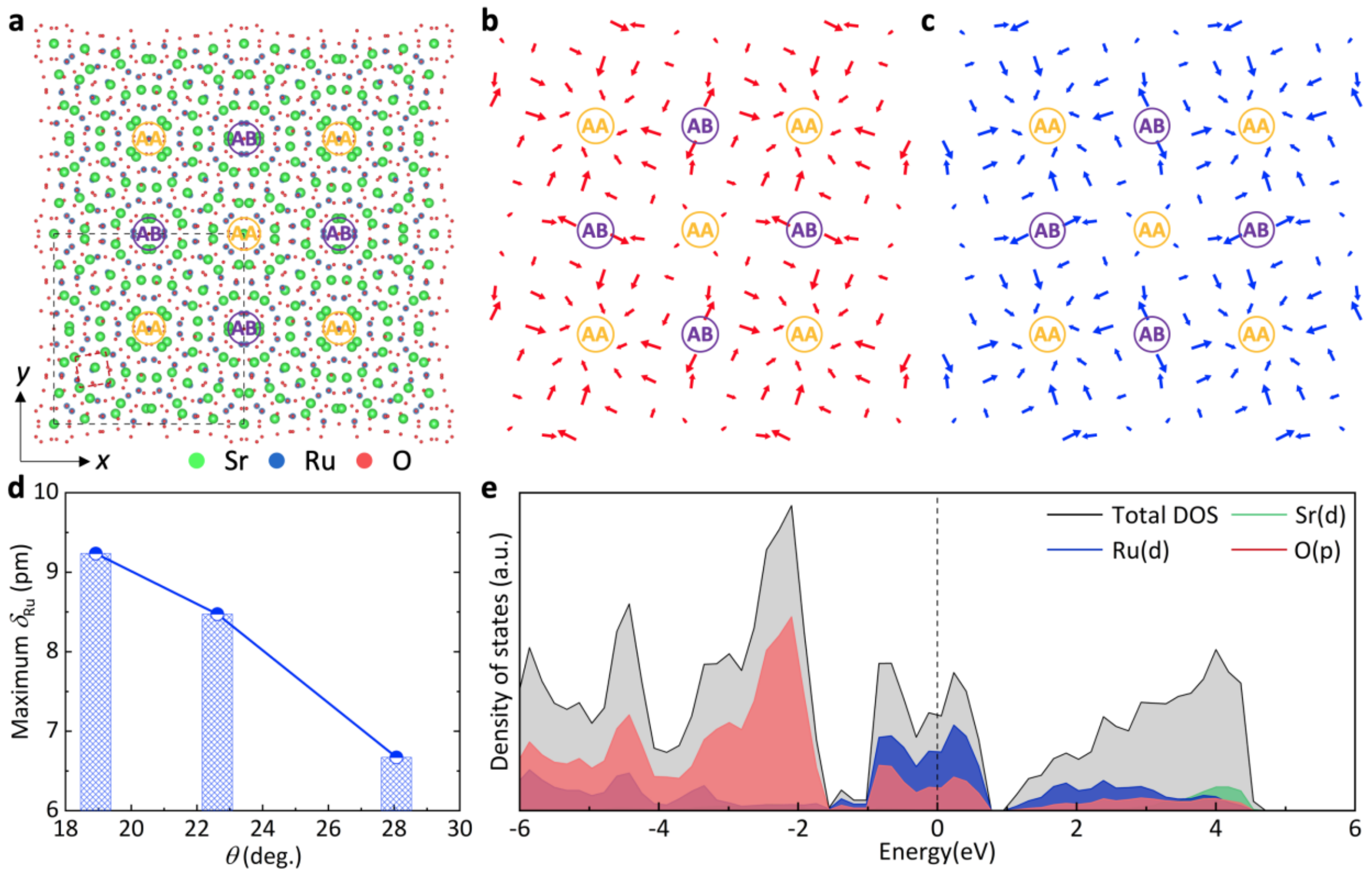

**Figure 3. DFT calculation for polar vortices in metallic SRO.** **a**, Planar view of the constructed 2x2 supercell for the 18.92° twisted bilayer. The black and red dashed squares denote the supercell and unit cell of SRO, respectively. **b-c**, In-plane Ru displacement maps of the top (red) and bottom (blue) layers, respectively. All displacements are amplified by a factor of 25 for clarity. The AA- and AB-stacked regions are marked with yellow and purple circles, respectively. **d**, Maximum in-plane Ru displacements as functions of the twist angle, extracted from the calculations. **e**, Electronic density of states of the polarized twisted bilayer. The Fermi level is indicated by a dashed line.

### Functional consequences of polar vortices

Having established the presence of polarization vortices in t-BL of metallic SRO, a natural question arises: is there any tangible impact on the functional properties? A first consideration is that bulk SRO is paramagnetic at room temperature, but ferromagnetic below ~160 K[31] (~150K for epitaxial films[32]). This ferromagnetic ordering manifests as a distinct change in electrical resistance[33,34]. We therefore measured the temperature-dependent magnetization (*M-T*), ferromagnetic hysteresis loop (*M-H*) and resistance as a function of temperature (R-T) for SL and t-BL SROs, as shown in Figs. 4a-c respectively (measurement details in Methods). The Curie temperature $T_C$ (inset of Fig. 4a) and the saturation magnetization $M_s$ (hysteresis loops in Fig. 4b) decrease as the twist angle is reduced. Meanwhile, figure 4c confirms that all samples exhibit metallic conductivity, and a kink in resistance due to the ferromagnetic transition shows the same twist angle-dependence as the magnetization. These results thus indicate the coexistence of polar vortices and ferromagnetic ordering, so that SRO is a “multiferroic metal” under twisted stacking.

The angle dependence of the magnetization is nevertheless counter-intuitive, since one might expect $T_c$ and magnetization to converge towards the SL value as the angle approaches 0°, yet at low angles we see the largest difference. We further observe that, when the twist-dependent $T_C$ and $M_s$ are compared with $\delta_{Ru}$ (Fig. 4d), the polarization and the ferromagnetism are anticorrelated, echoing the known difficulty of stabilizing ferroelectricity and ferromagnetism within the same material[35,36]. To capture the microscopic mechanism of this anticorrelation, we adopted a localized model based on the SRO bulk orthorhombic structure, in which Ru atoms were then manually off-centered (directly simulating the magnetism of realistic small-angle twisted bilayers using DFT would be computationally impractical because the corresponding moiré supercells contain thousands of atoms). The magnetic exchange parameter of the Heisenberg model was then calculated based on DFT total energies of ferromagnetic and anti-ferromagnetic phases (see Methods). The results, shown in Fig. 4e reveal a gradual decrease in the exchange parameter with increasing polarization, consistent with the observed lower magnetization. An explanation thus emerges: flexoelectrically-induced Ru displacement results in a geometric tilting of the Ru-O-Ru bond angle and thus a decrease in the magnetic exchange. The Curie temperature estimated based on the tilt-dependent exchange parameter therefore decreases with increasing polar distortion, in agreement with the experimental observations (Fig. 4f).

We have examined other possibilities. Cross-sectional STEM (Extended Data Fig. 1) does not show interfacial reconstruction or chemical diffusion, which are thereby excluded as primary causes of magnetic suppression. Meanwhile, misalignment between the magnetic easy axes of the two layers also could influence the net magnetization, but such effect would increase magnetic frustration at larger twist angles, which is opposite to our experimental trend and thus also excluded.

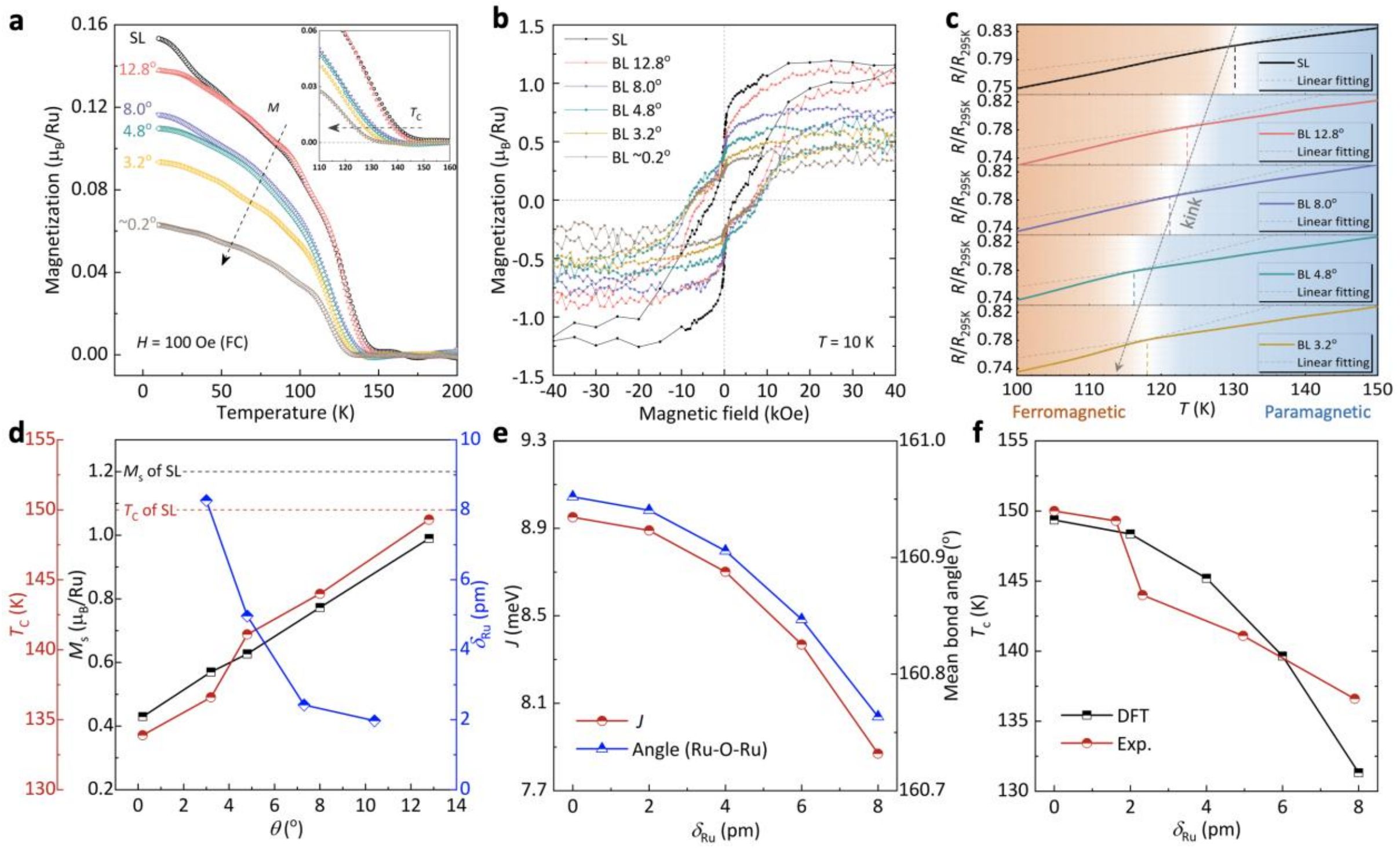

**Figure 4. Magnetic and electrical transport characterizations of polarized t-BL $SrRuO_3$.** **a**, Temperature-dependent magnetization field-cooling curves of SL and BL SROs measured under an in-plane magnetic field of 100 Oe. The inset highlights the variations in ferromagnetic transition temperature. **b**, In-plane ferromagnetic loops measured at 10 K. **c**, Normalized temperature-dependent resistance curves. The vertical dashed lines indicate the kinks in resistance, which is determined by linearly fitting the local raw data on the right and left sides. **d**, Twist-angle dependencies of ferromagnetic transition temperature ($T_C$), saturation magnetization ($M_s$) and Ru displacement ($\delta_{Ru}$). The dashed lines indicate the $T_C$ and $M_s$ of SL SRO. The mean values of $\delta_{Ru}$ are extracted from the Gaussian fitting in Fig. 2i. **e**, Magnetic exchange parameter ($J$) and mean Ru-O-Ru bond angle as a function of Ru displacement. **f**, Comparison of $T_C$ from DFT calculations and experiments. The experimental values were obtained via linear interpolation based on the mean $\delta_{Ru}$ and $T_C$ across samples with varying twist angles.

Direct measurement of magnetoelectric coupling is difficult in a metal, as any externally applied voltage would be screened by free carriers, but there might be indirect strategies. For example, adding an electromechanical element into the stack would allow voltage to generate strain which, in turn, would modulate the moiré configuration. Conversely, magnetostriction in SRO may also alter the strain and hence the polarization. Detecting these polarization changes will nevertheless remain challenging as long as we rely on in-situ STEM measurements. Moving forward, it would be useful to explore alternative approaches, such as optical second-harmonic generation[19], to detect polarization. The small lateral size of the vortices, however, remains a challenge for these alternatives.

## Conclusion

These findings expand the field of polar topology[37,38] into metals and strongly correlated electron systems[39,40], where it can have spintronic consequences such as the flexo-Rashba effect[20] or topologicall Hall effects. . Since flexoelectricity is a universal property, we expect similar phenomena to appear in twisted bilayer metallic systems beyond SRO. Furthermore, these findings are established in the experimentally available intermediate-angle regime and are observed at mesoscopic (> 3 nm) distances from the interface; the increasing moiré length scale and possible structural reconstructions or dislocations in bilayers with tiny twist angles may introduce both bigger lateral scale and finer thickness scale that warrant further investigation.

## Methods

### Epitaxial film growth

All epitaxial films were synthesized using pulsed-laser deposition (PLD) with a KrF excimer laser (248 nm, COMPex 102, Lambda Physik). The $SrRuO_3$ thin films were epitaxially grown on unetched (100)-oriented $SrTiO_3$ substrates with an 8.8 nm-thick $Sr_3Al_2O_6$ sacrificial layer to enable the subsequent membrane release. Prior to deposition, the chamber was evacuated to a base pressure of $1.0 \times 10^{-4}$ mTorr. Stoichiometric ceramic targets were used to ensure the chemical composition of films. The SAO layer was deposited at 750 °C under an oxygen pressure of 1.5 mTorr, with a laser fluence of 1.60 J/cm$^2$. The SRO layer was subsequently grown at 700 °C under an oxygen pressure of 100 mTorr $O_2$, with a laser fluence of 1.28 J/cm$^2$. The laser repetition rate was maintained at 1 Hz throughout the deposition process. After deposition, the films were cooled down to room temperature at 5 °C/min under an oxygen pressure of 10 Torr.

### Twisted bilayer membrane transfer

To obtain freestanding SRO membranes, the epitaxial films were adhered to a polydimethylsiloxane (PDMS) stamp and immersed in deionized water to dissolve the SAO sacrificial layer. The complete dissolution of SAO led to the detachment of the film from the substrate. The released membrane, attached to the PDMS stamp, was then transferred onto a target substrate using a transfer platform that allows to control the rotated angle of the substrate using optical alignment. Half of the membrane on the PDMS was brought into contact with the substrate, and the temperature was then increased to 70 °C to remove residual water. The PDMS was subsequently peeled off slowly, ensuring an intact transfer of the membrane. To make a twisted-bilayer sample, the substrate with the transferred bottom layer was rotated to a predefined angle using a micrometer probe in the transfer system. The remaining half of the membrane on the PDMS was then transferred onto the bottom one in the same manner, thereby obtaining twisted-bilayer samples. To enhance interlayer interaction and eliminate possible adsorbates at the interface, the samples were annealed at 330 °C for 8 hours under oxygen flow rate of 50 sccm.

### X-ray diffraction (XRD) and surface characterizations

High-resolution $2\theta$-$\omega$ scans and reciprocal space mappings (RSMs) were conducted using a Panalytical X'pert Pro diffractometer with a Cu K$\alpha_1$ (1.540598Å) source. The system featured a hybrid two-bounce primary monochromator on the incident beam side and a PIXcel line detector for high-resolution data acquisition. The surface morphology was characterized using MFP-3D Asylum AFM (Asylum Research -Oxford Instruments). Tapping mode was performed to minimize surface damage using a soft tip.

**TEM, SAED, and STEM characterizations**

The low-magnification HRTEM images and SAED patterns were acquired using a FEI Tecnai F20 TEM at 200 kV. High-quality cross-sectional samples were fabricated using the focused ion beam (FIB) technique at a Thermo Fisher Helios 5 UX dual-beam system. The STEM experiments were conducted using a double aberration-corrected Spectra 300 operated at 300 kV. Depth-sectioning STEM-HAADF was performed by acquiring a series of atomic-resolution images at different defocus depth[8], allowing for discrimination between the top layer and the bilayer interface. STEM-HAADF images were acquired with a probe convergence semi-angle of 27.3 mrad while the HAADF detector collection semi-angle was set as 63–200 mrad. ABF-STEM image was acquired with a collection semi angle range of 3-15 mrad.

**Polarization vortices and strain analysis**

The atomic positions of Ru and Sr were firstly extracted from Wiener-filtered STEM-HAADF images using the Python package Atomap[41]. After locating the Ru and Sr columns, the off-center displacement of each unit cell was calculated as the deviation of the Ru column from the mass center of the four adjacent Sr columns. The mean Ru displacement value is subtracted in the mappings to improve the visibility of the polarization vortex array. The polarization vortices of Ru with respect to the oxygen columns were extracted from the ptychographic reconstruction using the same method. Strain analysis was performed using the Geometrical Phase Analysis (GPA)[42] software package (HREM Research) in Digital Micrograph.

**Ptychographic STEM reconstruction**

Four-dimensional (4D) STEM datasets for multislice electron ptychography were acquired using an electron microscope pixel array detector (EMPAD). The probe convergence semi-angle was set to 28.4 mrad. The probe was overfocused by approximately 10 nm relative to the top surface of the sample and scanned over a raster grid with a step size of 0.52 Å. Each diffraction pattern consisted of 128 × 128 pixels with a reciprocal-space pixel size of 0.055 $Å^{-1}$. The dataset was acquired with a beam current of approximately 40 pA and a dwell time of 1 ms per probe position, resulting in a total electron dose of ~$9 \times 10^5$ $e^-$ $Å^{-2}$. Optimal parameters for the ptychographic reconstruction were determined using a Bayesian optimization framework[43] implemented in the *fold-slice* package[28,29]. Ptychographic reconstructions were performed using Ptyrad[30]. During reconstruction, the diffraction patterns were padded to 160 × 160 pixels, corresponding to a real-space pixel size of 0.09 Å. The reconstructions were run for 1100 iterations, employing a slice thickness of 0.5 nm and six probe modes.

**Electrical transport and magnetization measurements**

Temperature-dependent resistance measurements were performed in an Advanced Research Systems cryostat using a Keithley 2614B SourceMeter. The SL and t-BL membranes were transferred onto an insulating oxidized silicon substrate, and the Pt electrode arrays were sputtered through a square array-mask placed on the sample surface, defining 100 μm electrodes separated by 20 μm spacing. The electrodes were contacted via wire-bonding to a leadless chip carrier (Kyocera) electrically connected to the sourcementer by a chip-carrier socket, and thermally connected to the cold finger of the cryostat. The resistance was measured using two-point contact mode, as the temperature gradually varied at a rate of 1 K/min. Ferromagnetic hysteresis loops and temperature-dependent magnetization field-cooling curves were measured using a vibrating sample magnetometer (VSM, Quantum Design MPMS-3). The SL and t-BL membranes were transferred onto sapphire substrates. The external magnetic field was applied parallel to the membrane plane.

**DFT calculation for polar vortex**

DFT calculations were performed using the Vienna Ab Initio Simulation Package (VASP)[44,45]. The projector-augmented wave (PAW) method[46] was employed, with the Perdew–Burke–Ernzerhof (PBE)[47] generalized gradient approximation (GGA)[48] used as the exchange-correlation functional. The cubic structure was adopted to construct bilayer supercell structures, following the method proposed by Lee et al.[9]. The electronic wave functions were expanded using a plane-wave basis set with a kinetic energy cutoff of 600 eV and the 12×12×1 k-point mesh was used to sample the Brillouin zone. The monolayer structure was built based on the relaxed bulk SRO and was fully relaxed. The equilibrium lattice constant of cubic monolayer was calculated to be 3.781 Å, and the same value was used for all bilayer structures. The 12×12×1 k-mesh was used for monolayer and untwisted bilayer structures, whereas Γ-point sampling was adopted for the twisted structures. Structural relaxations were performed until the Hellmann-Feynman forces on each atom were reduced below 0.01 eV/Å. Van der Waals interactions between adjacent layers were accounted for using the Grimme D3 method[49]. To prevent spurious interlayer interactions, a vacuum region exceeding 15 Å was introduced in the supercell.

**DFT calculation for magnetism**

To examine the microscopic origin of magnetism suppression in twisted bilayer, density-functional theory (DFT) calculations were performed based on an orthorhombic structure with manually off-centering Ru along the *c*-axis direction to mimic the polarization observed in twisted bilayers. This localized modelling strategy was adopted because direct simulation of the experimentally relevant small-angle twisted bilayers is computationally impractical as the corresponding moiré supercells contain thousands to tens of thousands of atoms once commensurability, full structural relaxation, spin-orbit coupling, and magnetic self-consistency are considered.

The magnetism in twisted bilayer is primarily affected by twist-induced strain modulation. The elastic strain field generated at the interface propagates continuously through the entire membrane due to lattice continuity, resulting in local distortions in the internal lattice that share similar characteristics with those at the interface, differing mainly in magnitude. Within this framework, this localized modelling provides a physically reasonable approximation to evaluate how local structural distortions influence the magnetic exchange at both the interface and the strain-transferred internal regions.

In these calculations, a plane-wave basis set with a kinetic energy cutoff of 500 eV was used to expand the electronic wave functions, and an 8 × 8 × 6 k-mesh was employed to sample the Brillouin zone. The on-site Coulomb interaction parameters were set to U = 3 eV and J = 0.3 eV for Ru 4d orbitals. The Curie temperature was estimated using the Heisenberg model and the self-consistent Gaussian approximation (SCGA) method[50]. Specifically, the magnetic exchange parameter ($J$) of Heisenberg model was derived from the DFT calculated total energies of ferromagnetic and G-type antiferromagnetic phases[51,52], as shown in Supplementary Fig. S13,

$$J = (E_{\mathrm{AFM}} - E_{\mathrm{FM}}) / (N_{\mathrm{nn}} \cdot N_{\mathrm{mag}}), \tag{1}$$

where $E_{\mathrm{AFM}}$ ($E_{\mathrm{FM}}$) is the total energy of the ferromagnetic and antiferromagnetic phases, $N_{\mathrm{mag}}$ = 4 is the number of magnetic Ru ions in the unit cell, and $N_{\mathrm{nn}}$ = 6 is the number of nearest-neighbour Ru ions of each Ru ion. The Curie temperature was estimated using SCGA method[50] as

$$T_C^{SCGA} \simeq \frac{1}{3}\theta_C N_{\mathrm{nn}} J, \tag{2}$$

$\theta_C$ = 0.719 is a structure dependent parameter[51].

## Acknowledgments

The authors acknowledge support by the National Natural Science Foundation of China (Grant Nos. 12402183, 12374080), the National Key Projects for Research and Development of China (Grant No. 2021YFA1400300), and the Beijing Natural Science Foundation (Grant No. 1244057). We also acknowledge ICN2 and the Joint Electron Microscopy Center at ALBA (JEMCA) for providing key facilities and technical guidance. ICN2 is supported by the Severo Ochoa program from Spanish MCIN / AEI (Grant No. CEX2021-001214-S), the European Regional Development Fund from European Union (Grant No. IU16-014206 (METCAM-FIB)), and the CERCA Programme, Generalitat de Catalunya. ICN2 is a founding member of e-DREAM. J.A. and X.H. acknowledge funding from Generalitat de Catalunya (Grant No. 2021SGR00457). This study is part of the Advanced Materials programme and was supported by MCIN with funding from European Union NextGenerationEU (PRTR-C17.I1) and by Generalitat de Catalunya (In-CAEM Project). We acknowledge support from CSIC Interdisciplinary Thematic Platform (PTI+) on Quantum Technologies (PTI-QTEP+). This work has been funded by the European Commission – NextGenerationEU (Regulation EU 2020/2094), through CSIC's Quantum Technologies Platform (QTEP). ICN2 acknowledges funding from Grant IU16-014206 (METCAM-FIB) funded by the European Union through the European Regional Development Fund (ERDF), with the support of the Ministry of Research and Universities, Generalitat de Catalunya. Y.L. acknowledges support from the Severo Ochoa Seed Funding program (Grant CEX2021-001214-S/MICIU/AEI/10.13039/501100011033). D.P. acknowledges the "Consolidación Investigadora" grant CNS2024-154522 and grant PID2022-140589NB-I00, funded by MCIN/AEI/10.13039/501100011033. This work is in the framework of the Universitat Autonoma de Barcelona Materials Science PhD program. X.H. acknowledges PhD scholarship support from the China Scholarship Council (CSC) (Grant No. 202304910019). Q. R. acknowledges support from the China Scholarship Council (CSC) (Grant No. 202506030153), and the Postdoctoral Fellowship Program (Grade B) of China Postdoctoral Science Foundation (Grant No. GZB20250965). U.S. acknowledges the funding for this project by Grant PID2019-108573GB-C21 (FOx-Me) funded by the Spanish Ministry of Science and Innovation (DOI: 10.13039/501100011033). I.P.-H. acknowledges funding from AGAUR-FI scholarship (Grant No. 2023FI-00268) Joan Oró of the Secretariat of Universities of the Generalitat of Catalonia and the European SocialPlus Fund. Calculations were performed using resources of the Supercomputer Centre in Chongqing. The authors acknowledge the use of the Spectra 300 microscope, provided under ALBA Synchrotron proposal (Grant No. 20240320035). The authors gratefully acknowledge the assistance of Francesco Salutari, Athique Ahmed Ali and Francisco Javier Belarre Triviño in the preparation of cross-sectional FIB samples, and Bernat Bozzo in the magnetic measurements, and Riccardo Rurali in DFT calculations.

Note: after writing this manuscript, we have become aware of another work on $SrRuO_3$ membranes (Alejandro Martin Merodio et al., group of Jacobo Santamaría at the Universidad Complutense de Madrid, presented at QUOROM, 22/05/25). So far, they only report transport properties, but we expect their membranes will also display vortices.

**Author contributions**

Experimental measurements were mainly performed at ICN2 and the Joint Electron Microscopy Center at ALBA (JEMCA), and theoretical calculations were carried out at BIT. All institutions contributed substantially to the development and completion of this work. G.C., J.H., and J.A. designed and supervised the project. Y.L. conceived the original idea for this work. Y.L. fabricated the thin films and transferred the samples with the help from U.S. and J.M.C.R.. X.H., J.A., B.M., K.G., and I.P. conducted STEM, SAED and TEM experiments. X.H. and Y.L. performed data analysis on the polar vortices, strain gradients, and flexoelectric fields. Q.R., B.Z., G.T., and J.H. carried out the atomistic simulations and analysis. Y.L. conducted the ferromagnetism and resistance measurements under the guidance of D.P. and X.W.. Y.L. performed the surface morphology measurements. J.P. and J.S. conducted the XRD measurements. Y.M. and Q.L. provided assistance with the experiments. M.B.E and G.M provided assistance with the ptychography reconstruction. Y.L., X.H. and G.C. wrote the manuscript with the help from all authors. All authors contributed to the discussion of the results and the revision of the manuscript.

**Competing interests**

The authors declare that they have no competing interests.

**Data and materials availability**

All data are available in the manuscript or the supplementary information.

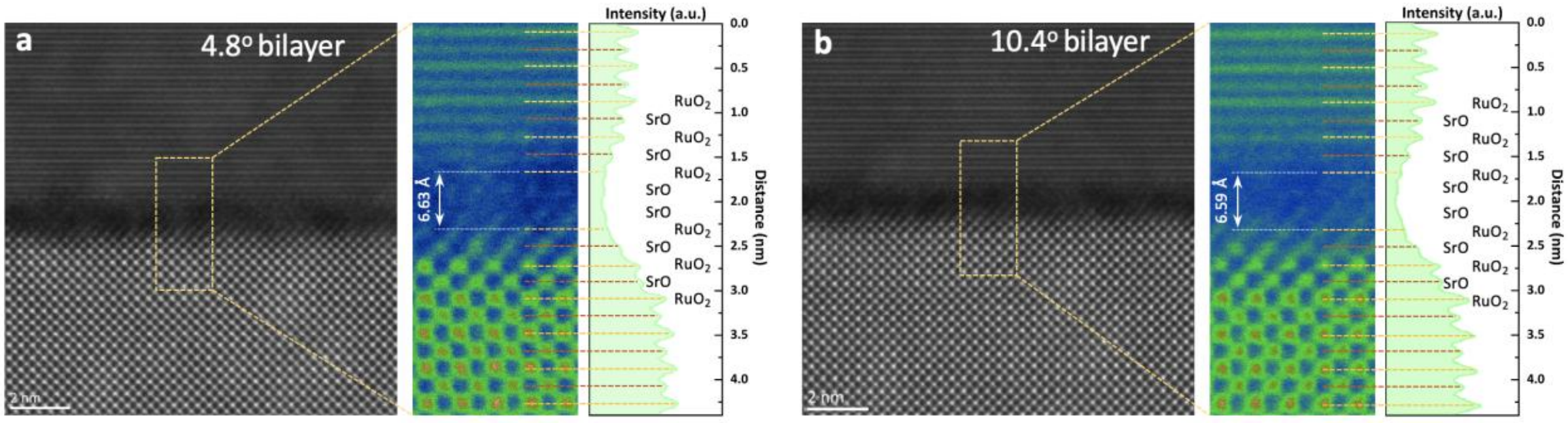

**Extended Data Figure 1.** Atomic resolution cross-sectional HAADF-STEM characterizations for interface of twisted-bilayers. **a-b**, Transferred 4.8° and 10.4° twisted bilayer with bottom-layer-aligned zone axis, respectively. The interface structural characteristics are the same for both bilayers with large and small angles, as expected from their identical fabrication conditions. Left panels show the HAADF-STEM image of the atomic interface, in which periodic and parallel (001) lattice fringes and atomic column are found in the both layers. The (001) planes on the top and bottom layers are parallel, as required for, and expected from a coherent bilayer structure. The atomic column intensity decays towards the interface, yielding a dark region due to the low Z-contrast scattering of SrO termination. The magnified area at around the interface (middle panel) shows the atomic columns more clearly and allows further analysis. The corresponding intensity line profile across the interface shows that the last two visible $RuO_2$ planes (indicated by orange dash lines) are separated by a distance larger than the bulk lattice constant, at ~6.6 Å (right panel). This, however, is a conservatively estimated distance and unlikely to be the true interlayer distance, as atomic columns and planes are still distinguishable inside the interfacial gap.

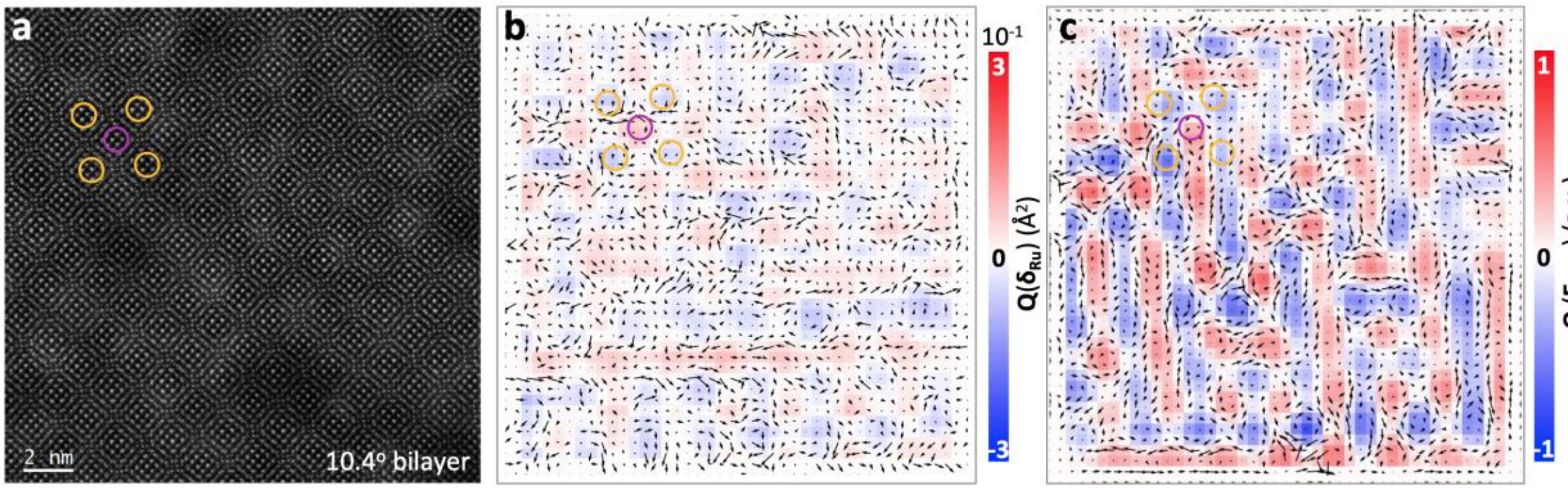

**Extended Data Figure 2. a**, Planar-view STEM-HAADF images focusing on the interface in the 10.4° twisted bilayer, showing distinct moiré patterns. **b**, Ru displacement maps of the top layer superimposed with their toroidal moment $Q(\delta_{Ru})$. The Ru displacements are amplified by a factor of 120 for clarity. **c**, Normalized flexoelectric field vector maps induced by shear strain gradients in the top layer, superimposed with their toroidal moment $Q(E_{flexo})$. The yellow and purple circles indicate the AA- and AB-stacked regions, respectively.

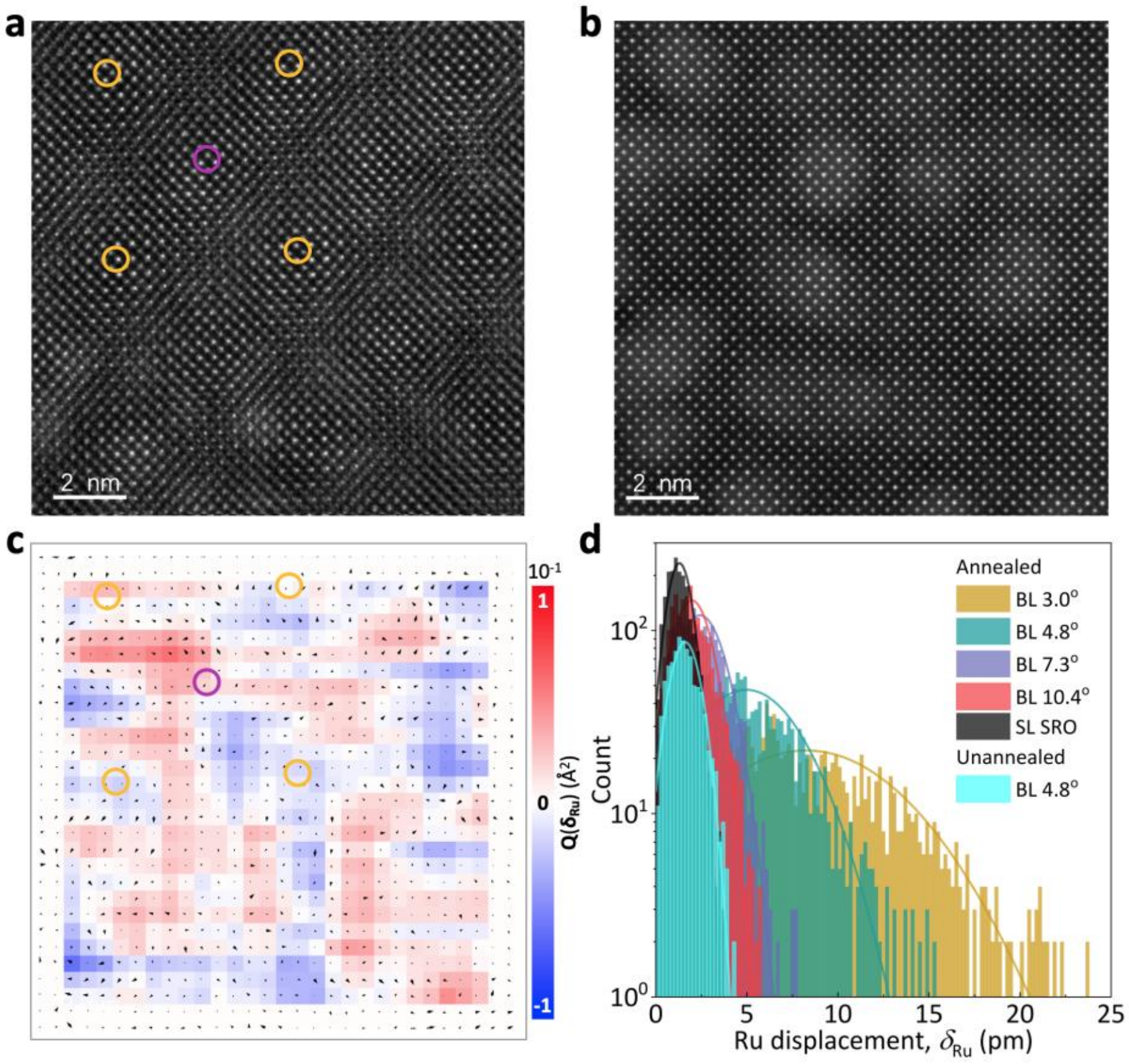

**Extended Data Figure 3.** Planar-view STEM-HAADF images of the 4.8° twisted bilayer without annealing, focusing on **a**, the bilayer interface and **b**, the upper surface of the top layer, respectively. **c**, Ru displacement map of the top layer superimposed with the toroidal moment (*Q*). All Ru displacements are amplified by a factor of 50 (same factor used for the annealed 4.8° twisted bilayer). The yellow and purple circles in **a** and **c** indicate the AA- and AB-stackings. **d**, Histograms comparing the magnitude of Ru displacement in annealed and unannealed samples. As shown in **c** and **d**, the Ru displacements in the unannealed samples are both random and minimal (1.6 pm on average), suggesting a nonpolarized state similar to that of SL SRO.

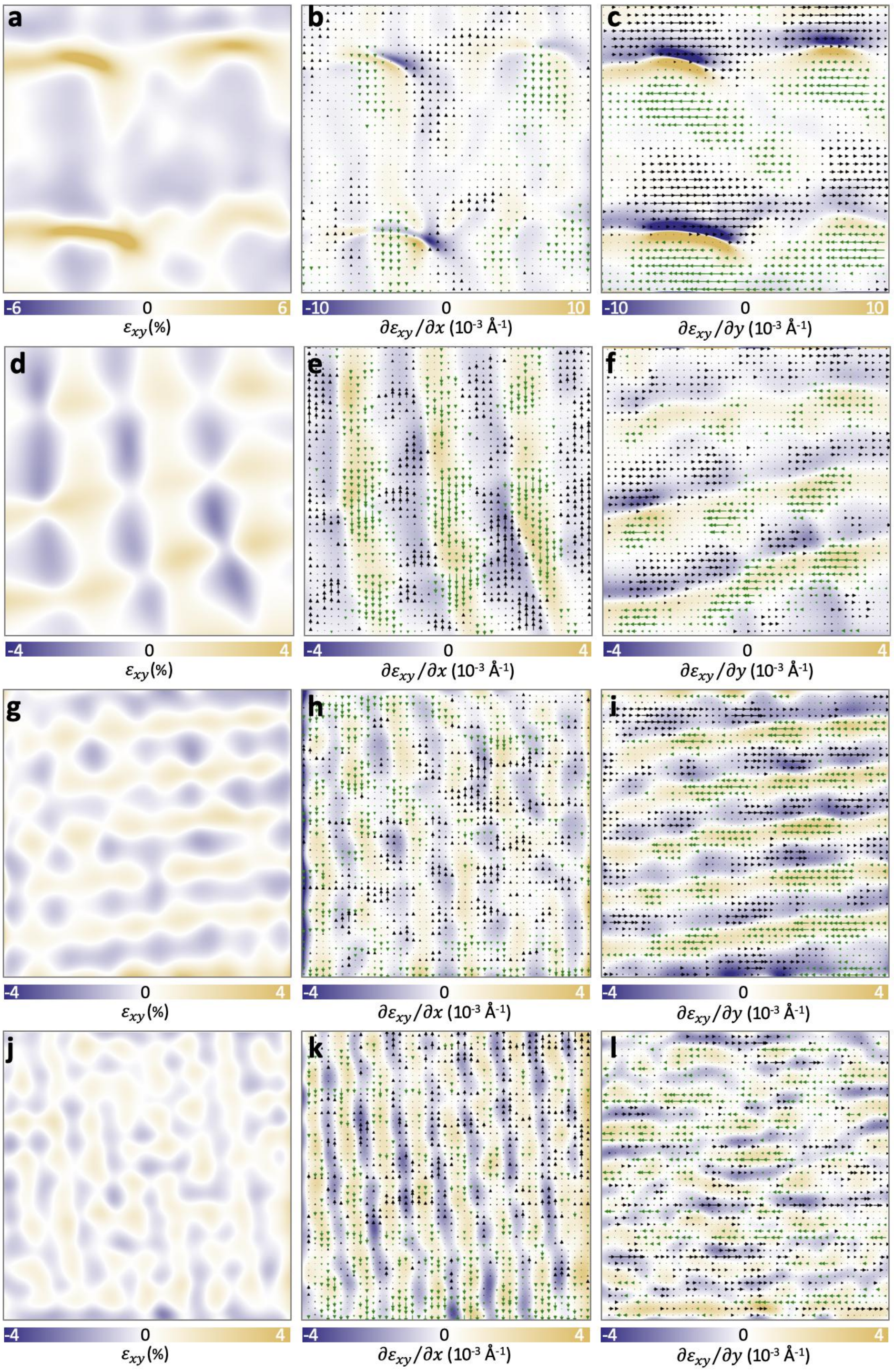

**Extended Data Figure 4. a**, Shear strain ($\varepsilon_{xy}$) component map for the top layer in 3.0° twisted bilayer. **b-c**, Corresponding shear strain gradient maps along the *x*-axis and *y*-axis, respectively, superimposed with Ru displacement components along the *y*-axis and *x*-axis. The strain gradients ($\varepsilon_{xy,x}$ and $\varepsilon_{xy,y}$) can generate flexoelectric field components in the *y*-direction and *x*-direction, respectively. Same shear strain (gradients) analyses for **d-f**, the 4.8° twisted, **g-i**, the 7.4° twisted and **j-l**, the 10.4° twisted bilayers. The Ru displacement components in **b-c**, **e-f**, **g-i**, and **j-l** are amplified by a factor of 30, 50, 100, and 120 for clarity, respectively.

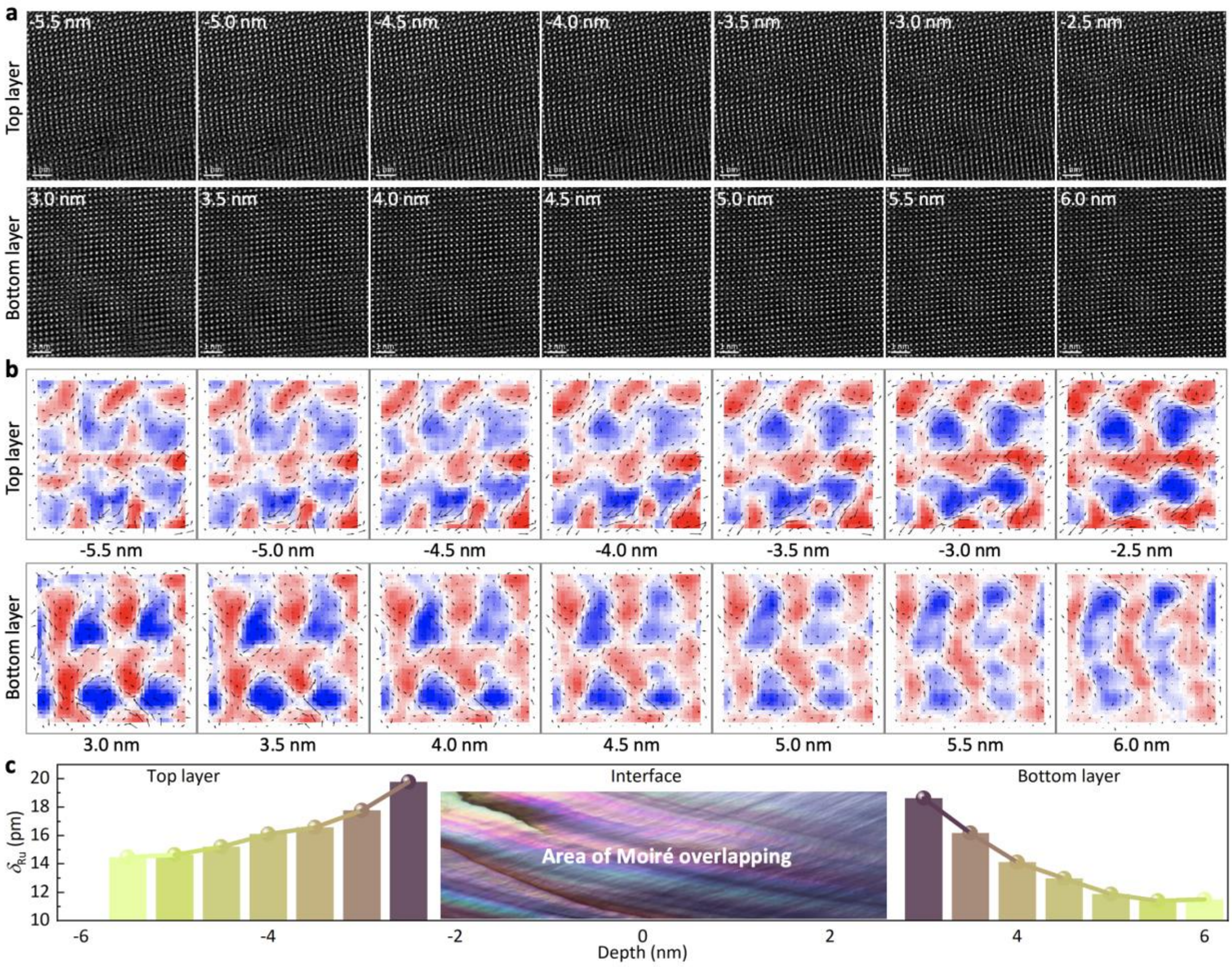

**Extended Data Figure 5**. **a**, Multislice electron ptychographic reconstruction of depth-resolved lattice structures in a 4.8° twisted bilayer. **b**. Corresponding Ru displacement vector maps superimposed with their toroidal moment at different depths. **c**, Depth-dependent mean Ru displacement extracted from Gaussian fitting of the displacement histograms in **b**. The presence of moiré interference fringes near the interface hinders accurate identification of atomic columns, making it challenging to reliably extract the polarization vector distribution and its magnitude. Depth denotes the distance from the interface, where negative and positive values correspond to the top and bottom layers, respectively. The variation of the reconstructed lattice and the corresponding polarization vortices as a function of depth can be found in Supplementary Movie S1.